# A New Formula for the BER of Binary Modulations with Dual-Branch Selection over Generalized-*K* Composite Fading Channels

Imran Shafique Ansari*, Saad Al-Ahmadi**, Ferkan Yilmaz*, Mohamed-Slim Alouini*, and Halim Yanikomeroglu***

*Electrical Engineering Program, Division of Physical Sciences and Engineering

King Abdullah University of Science and Technology, Thuwal, Makkah Province, Saudi Arabia

Email: {imran.ansari, ferkan.yilmaz, slim.alouini}@kaust.edu.sa

**Department of Electrical Engineering

King Fahd University of Petroleum and Minerals, Dhahran, Eastern Province, Saudi Arabia

Email: ahmadisa@kfupm.edu.sa

***Department of Systems and Computer Engineering

Carleton University, Ottawa, Ontario Province, Canada

Email: halim@sce.carleton.ca

**Abstract**

Error performance is one of the main performance measures and derivation of its closed-form expression has proved to be quite involved for certain systems. In this letter, a unified closed-form expression, applicable to different binary modulation schemes, for the bit error rate of dual-branch selection diversity based systems undergoing independent but not necessarily identically distributed generalized-$K$ fading is derived in terms of the extended generalized bivariate Meijer $G$-function.

**Index Terms**

Diversity schemes, selection combining, dual-branch selection diversity, binary modulation schemes, generalized-$K$ (GK) model, composite fading, bit error rate (BER), and Meijer $G$-function distribution.

## I. INTRODUCTION

In recent times, different diversity schemes have taken up an important position in the wireless communication systems. The main reason behind this is that these different diversity schemes allow for multiple transmission and/or reception paths for the same signal [1]. One of the simplest diversity combining scheme is the selection combining (SC) diversity scheme where only one of the diversity branches is processed. Specifically, SC scheme chooses the branch with highest signal-to-noise ratio (SNR) [2]-[3].

Additionally, wireless communications are driven by a complicated phenomenon known as radio-wave propagation that is characterized by various effects including multipath fading and shadowing. The statistical behavior of these effects is described by different models depending on the nature of the communication environment. It is becoming necessary to study such effects i.e. large-scale fading as well as small-scale fading concurrently as the multihop relay networks are emerging in the current times. Using the Nakagami multipath fading model that is versatile enough to model various multipath fading conditions ranging from severe fading to non-fading scenario, and the Gamma model for shadowing has led to the generalized-$K$ (Gamma-Gamma) composite fading model [4]-[8]. Generalized-$K$ (GK) distribution, earlier used in radar applications and recently being used in the context of wireless digital communications over fading channels, is one of the relatively new tractable models used to describe the statistical behavior of multipath fading and shadowing effects as compared to log-normal based models. GK fading model is quite general model as it includes *K*-distribution as its special case and accurately approximates many other fading models such as Nakagami-*m* and Rayleigh-Lognormal (R-L) ([6] and references therein). Finally, GK distribution is a distribution of the product of two independent Gamma random variables (RV) and hence is a special case of the Fox *H*-function and in turn a special case of Meijer *G*-function where the product of two Meijer *G*-function's can be represented in terms of a extended generalized bivariate Meijer *G*-function (EGBMGF) (refer Table I) [9].

It is noteworthy to mention that bit error rate (BER) is one of the most important performance measures that forms the basis in designing wireless communication systems. Based on the open technical literature and upto the best of our knowledge, error analysis has been performed for dual diversity with SC over log-normal fading channels in closed-form using moment generating function (MGF) based approach in [10] and with Weibull fading channel as an approximate using characteristic function (CF) based approach in [11]. Additionally, error performance analysis of SC systems with independent and identically distributed (i.i.d.) GK fading branches was performed in [12] involving integral form expressions. Further, in [13] the analysis was performed for dual-branch SC citing the difficulty in deriving the expression for the probability density function (PDF). This issue was tackled in [14] for an arbitrary number of branches and the authors therein have described and utilized a method to perform the BER analysis directly from the cumulative density function (CDF) eliminating the need of deriving the PDF and relying on the Gauss-Laguerre quadrature technique.

In this work, we revisit this problem under the umbrella of the *H*-functions and derive exact closed-form expression of the BER of binary modulation systems with dual-branch SC scheme and undergoing GK fading where the channels are independent but not necessarily identically distributed (i.n.i.d.). The remainder of the paper is organized as follows. Section II introduces the system and the GK channel model. Next, section III presents some statistical characteristics of GK fading channel model followed by the analytical BER analysis in section IV, and finally,

December 17, 2010                                                                                                                                              DRAFT

section V discusses the results followed by the summary of the paper in section VI.

## II. THE GENERALIZED-$K$ FADING SYSTEM AND CHANNEL MODEL

A SC based communication system with a source and a destination is considered with i.n.i.d. channels as follows

$$Y = \alpha X + n, \tag{1}$$

where $Y$ is the received signal at the receiver end, $X$ is the transmitted signal, $\alpha$ is the channel gain, and $n$ is the additive white Gaussian noise (AWGN). In a Nakagami multipath fading channel, $\gamma = |\alpha|^2$ follows Gamma distribution; additionally, the shadowing component is also assumed to follow a Gamma distribution. Hence, the channel gains experience composite fading whose statistics follow a generalized-$K$ distribution given by

$$p_\gamma(\gamma) = \frac{2b^{m_m+m_s}}{\Gamma(m_m)\Gamma(m_s)} \gamma^{\frac{m_m+m_s}{2}-1} K_{m_s-m_m}(2b\sqrt{\gamma}), \tag{2}$$

where $\Gamma(\cdot)$ is the Gamma function as defined in [15, Eq. (8.310)], $m_m$ and $m_s$ are the Nakagami multipath fading and shadowing parameters, respectively. In (2) $K_m(\cdot)$ is the modified Bessel function of the second kind and order $m$, $b = \sqrt{\frac{m_m m_s}{\Omega_0}}$, and $\Omega_0$ is the mean of the local power. The parameters $m_m$ and $m_s$ quantify the severity of multipath fading and shadowing, respectively, in the sense that small values of $m_m$ and $m_s$ indicate severe multipath fading and shadowing conditions respectively, and vice versa. The instantaneous SNR of the $n^{th}$ branch is given by $\gamma_n = (E_b/N_0)\, x_n^{\,2}$ where $x_n$ is the signal amplitude for the $n^{th}$ branch, $E_b$ is the average energy per bit and $N_0$ is the power spectral density of the AWGN.

## III. STATISTICAL CHARACTERISTICS

The PDF and CDF expressions of the GK RVs can also be written in terms of Meijer *G*-function.

*Lemma* 1: The PDF of a GK RV can be expressed in terms of Meijer-$G$ function as

$$p_\gamma(y) = \left(\frac{m_m m_s}{\Gamma(m_m)\Gamma(m_s)\Omega_0}\right) \times G^{2,0}_{0,2}\left[\left(\frac{m_m m_s}{\Omega_0}\right) y \,\middle|\, m_m-1, m_s-1\right], y > 0. \tag{3}$$

*Proof:* We may use the fact that the PDF of the product of $N$ independent Gamma RVs can expressed as a $H$-function PDF that is given by [16, Eq. (6.4.9)] as follows

$$p(x) = \left(\prod_{i=1}^{N} \frac{1}{\theta_i \Gamma(k_i)}\right) H_{0,N}^{N,0}\left[\left(\prod_{1}^{N} \frac{1}{\theta_i}\right) x \middle| (k_i - 1, 1), \ldots, (k_N - 1, 1)\right], x > 0. \quad (4)$$

Then, with $N = 2$, the GK PDF can be expressed as

$$p(x) = \left(\frac{m_m m_s}{\Gamma(m_m)\Gamma(m_s)\Omega_0}\right) H_{0,2}^{2,0}\left[\left(\frac{m_m m_s}{\Omega_0}\right) x \middle| (m_m - 1, 1), (m_s - 1, 1)\right], x > 0. \quad (5)$$

Now by applying [16, Eq. (6.2.8)], the expression in (3) follows. ∎

Further, substituting (3) in [17, Eq. (26)] and utilizing [16, Eq. (6.2.4)], the CDF of GK can be written as

$$P_\gamma(\gamma) = \frac{1}{\Gamma(m_m)\Gamma(m_s)} G_{1,3}^{2,1}\left[\left(\frac{m_m m_s}{\Omega_0}\right)\gamma \middle| \begin{array}{c} 1 \\ m_m, m_s, 0 \end{array}\right], \gamma > 0, \quad (6)$$

where $G[\cdot]$ is the Meijer $G$-function [15].

## IV. BER ANALYSIS

In SC combining scheme, the highest SNR branch is selected. In our case, for dual-diversity, the SNR, $\gamma_{sc}$, is given by

$$\gamma_{sc} = \max(\gamma_1, \gamma_2). \quad (7)$$

The CDF of $\gamma_{sc}$ is given by

$$F(\gamma_{sc}) = Pr(\max(\gamma_1, \gamma_2) \leq \gamma_{sc}) = \prod_{n=1}^{2} F_{\gamma_n}(\gamma_{sc}). \quad (8)$$

The BER for SC is given by

$$P_e = \int_0^\infty P_e(\epsilon|\gamma_{sc}) f_{\gamma_n}(\gamma) d\gamma_{sc} = \int_0^\infty P_e(\epsilon|\gamma_{sc}) dF_{\gamma_n}(\gamma_{sc}), \quad (9)$$

where $P_e(\epsilon|\gamma_{sc})$ is the conditional error probability (CEP) for the given SNR. A unified CEP expression for coherent and non-coherent binary modulation schemes over an AWGN channel is given in [18] as

$$P_e(\epsilon|\gamma_{sc}) = \frac{\Gamma(p, q\gamma_{sc})}{2\Gamma(p)}, \quad (10)$$

where $\Gamma(\cdot,\cdot)$ is the complementary incomplete gamma function [15, Eq. (8.350.2)]. The parameters $p$ and $q$ in (10) account for different modulation schemes. For an extensive list of modulation schemes represented by these parameters, one may look into [19]. Now, applying integration by parts in (9), we get

$$P_e = P_e\left(\epsilon|\gamma_{sc}\right)F(\gamma_{sc})\big|_0^\infty - \int_0^\infty F(\gamma_{sc})dP_e\left(\epsilon|\gamma_{sc}\right). \tag{11}$$

The first term goes to zero using [20, Eq. (6.5.3)]. Further, substituting (10) into (11) and using [20, Eq. (6.5.25)], the average BER can be written as

$$P_e = \frac{q^p}{2\Gamma(p)}\int_0^\infty \exp(-q\gamma_{sc})\gamma_{sc}^{p-1}F(\gamma_{sc})d\gamma_{sc}. \tag{12}$$

On substituting (8) in the above obtained expression, we get

$$P_e = \frac{q^p}{2\Gamma(p)}\int_0^\infty \exp(-q\gamma_{sc})\gamma_{sc}^{p-1}\prod_{n=1}^{2}F_{\gamma_n}(\gamma)d\gamma_{sc}. \tag{13}$$

Using [9], we obtain the product of the CDFs present in the above expression in terms of EGBMGF as

$$\prod_{n=1}^{2}F_{\gamma_n}(\gamma) = F_{\gamma_1}(\gamma)F_{\gamma_2}(\gamma) = \kappa_1\ S\left[\begin{array}{c}\left\{\begin{array}{c}\begin{bmatrix}0,0\\0,0\end{bmatrix}\\1,2\\0,1\\1,2\\0,1\end{array}\right\}\end{array}\middle|\begin{array}{c}-;-\\1;\kappa_2\\1;\kappa_3\end{array}\middle|\begin{array}{c}(\kappa_4)\,\gamma\\(\kappa_5)\,\gamma\end{array}\right], \tag{14}$$

where $S[\cdot]$ is the EGBMGF as given in [21, Eq. (2.1)], $\kappa_1 = \frac{1}{\Gamma(m_{m_1})\Gamma(m_{s_1})\Gamma(m_{m_2})\Gamma(m_{s_2})}$, $\kappa_2 = m_{m_1}, m_{s_1}, 0$, $\kappa_3 = m_{m_2}, m_{s_2}, 0$, $\kappa_4 = \frac{m_{m_1}m_{s_1}}{\Omega_{o_1}}$ and $\kappa_5 = \frac{m_{m_2}m_{s_2}}{\Omega_{o_2}}$. The above expression can also be expressed as,

$$\prod_{n=1}^{2}F_{\gamma_n}(\gamma) = F_{\gamma_1}(\gamma)F_{\gamma_2}(\gamma) = \kappa_1\ G_{0,0:1,3:1,3}^{0,0:2,1:2,1}\left(\begin{array}{c}-\\-\end{array}\middle|\begin{array}{c}1\\\kappa_2\end{array}\middle|\begin{array}{c}1\\\kappa_3\end{array}\middle|(\kappa_4\gamma),(\kappa_5\gamma)\right), \tag{15}$$

where $G(\cdot)$ is EGBMGF as in [22]. Additionally, (14) or (15) can be represented as,

$$\prod_{n=1}^{2} F_{\gamma_n}(\gamma) = F_{\gamma_1}(\gamma)F_{\gamma_2}(\gamma) = \kappa_1 \, S\left[\kappa_4\gamma, \kappa_5\gamma \,\middle|\, \begin{bmatrix} 0,0 \\ 0,0 \end{bmatrix} \,\middle|\, \begin{matrix} - \\ - \end{matrix} \,\middle|\, \begin{pmatrix} 2,1 \\ 1,3 \end{pmatrix} \,\middle|\, \begin{matrix} 1 \\ \kappa_2 \end{matrix} \,\middle|\, \begin{pmatrix} 2,1 \\ 1,3 \end{pmatrix} \,\middle|\, \begin{matrix} 1 \\ \kappa_3 \end{matrix} \right], \tag{16}$$

where $S[\cdot]$ is EGBMGF as in [23, Eq. (4)].

*Lemma* 2 [21, Eq. (2.1)]: The integral involving the EGBMGF of two variables with an exponential term with the RV as one of its argument and a term with RV itself evaluates to

$$\int_0^\infty x^{\lambda-1} e^{-\mu x} \, S\left[ \begin{bmatrix} p,0 \\ A-p, B \end{bmatrix} \begin{pmatrix} q,r \\ C-q, D-r \end{pmatrix} \begin{pmatrix} k,l \\ E-k, F-l \end{pmatrix} \,\middle|\, \begin{matrix} (a);(b) \\ (c);(d) \\ (e);(f) \end{matrix} \,\middle|\, \begin{matrix} \alpha x^\rho \\ \beta x^\rho \end{matrix} \right] dx \tag{17}$$

$$= (2\pi)^{\frac{1}{2}(1-\rho)} \frac{\rho^{\lambda-1/2}}{\mu^\lambda} \, S\left[ \begin{bmatrix} p+\rho,0 \\ A-p, B \end{bmatrix} \begin{pmatrix} q,r \\ C-q, D-r \end{pmatrix} \begin{pmatrix} k,l \\ E-k, F-l \end{pmatrix} \,\middle|\, \begin{matrix} \Delta(\rho,\lambda),(a);(b) \\ (c);(d) \\ (e);(f) \end{matrix} \,\middle|\, \begin{matrix} \frac{\alpha\rho^\rho}{\mu^\rho} \\ \frac{\beta\rho^\rho}{\mu^\rho} \end{matrix} \right],$$

where $\Delta(\rho, \lambda) = \frac{\lambda}{\rho}, \frac{\lambda+1}{\rho}, \cdots, \frac{\lambda+\rho-1}{\rho}$ [21, Eq. (1.7)].

Now, substituting (14) or (15) or (16) into (13), then using the Lemma 2 given above and performing additional manipulations, we get the desired closed-form expression for the average BER as

$$P_e = \frac{\kappa_1}{2\Gamma(p)} \, S\left[ \begin{bmatrix} 1,0 \\ 0,0 \end{bmatrix} \begin{pmatrix} 1,2 \\ 0,1 \end{pmatrix} \begin{pmatrix} 1,2 \\ 0,1 \end{pmatrix} \,\middle|\, \begin{matrix} p \\ 1;\kappa_2 \\ 1;\kappa_3 \end{matrix} \,\middle|\, \begin{matrix} \frac{(\kappa_4)}{q} \\ \frac{(\kappa_5)}{q} \end{matrix} \right], \tag{18}$$

or equivalently

$$P_e = \frac{\kappa_1}{2\Gamma(p)} \, G^{1,0:2,1:2,1}_{1,0:1,3:1,3} \left( p \, \left| \begin{array}{c} 1 \\ \kappa_2 \end{array} \right| \left. \begin{array}{c} 1 \\ \kappa_3 \end{array} \right| (\kappa_4)\frac{1}{q}, (\kappa_5)\frac{1}{q} \right), \tag{19}$$

or equivalently

$$P_e = \frac{\kappa_1}{2\Gamma(p)} \, S \left[ \frac{\kappa_4}{q}, \frac{\kappa_5}{q} \, \left| \begin{bmatrix} 1,0 \\ 1,0 \end{bmatrix} \right| p \, \left| \begin{pmatrix} 2,1 \\ 1,3 \end{pmatrix} \begin{array}{c} 1 \\ \kappa_2 \end{array} \right| \begin{pmatrix} 2,1 \\ 1,3 \end{pmatrix} \begin{array}{c} 1 \\ \kappa_3 \end{array} \right]. \tag{20}$$

## V. RESULTS AND DISCUSSION

The numerical results for BER of SC scheme with dual-diversity over i.n.i.d. GK fading channels are presented in this section.

The exact solution presented above in (18), (19) and/or (20) has not been found computable and hence its computability/evaluation was implemented using Mathematica as can be seen in Table II. With this implementation, the EGBMGF can be evaluated fast and accurately. An illustrative code is shown in Table III. This computability, therefore, has been utilized for different digital modulation schemes and is employed to discuss the results in comparison to respective Monte Carlo simulation outcomes.

The average SNR per bit in all the scenarios discussed is assumed to be equal. In addition, different digital modulation schemes are represented based on the values of $p$ and $q$ where $p = 0.5$ and $q = 1$ represents binary phase shift keying (BPSK), $p = 1$ and $q = 1$ represents differential phase shift keying (DPSK) and binary frequency shift keying (BFSK) is represented by $p = 0.5$ and $q = 0.5$. In Monte Carlo simulations, the GK fading channel was generated by the product of two independent gamma RVs.

We observe from figure 1 that this implemented computability of EGBMGF provides a perfect match to the MATLAB simulated results and the results are as expected i.e. the BER increases as the shadowing effect increases (i.e. value of $m_s$ decreases) while keeping multipath fading constant at $m_m = 1$. The figure shown represents BPSK. Its important to note here that these values for the parameters were selected randomly to prove the validity of the obtained results and hence specific values based on the standards can be used to obtain the required results.

Similar outcomes may be obtained for BFSK and DPSK. Additionally, similar analysis can be done for constant shadowing effect and varying the multipath fading.

Furthermore, to demonstrate the case that the results presented here also handle the presence of i.n.i.d. GK channels, following figure 2 presents the different modulation schemes with different effects of multipath fading and shadowing on both their channels. The values utilized for multipath fading and shadowing were as follows; $m_{m_1} = 1$, $m_{m_2} = 2$, $m_{s_1} = 0.5$, and $m_{s_2} = 4$. It can be seen that, as expected, BPSK outperforms the other modulation schemes and BFSK and DPSK perform in similar fashion at lower SNR whereas as the SNR increases DPSK performs better than BFSK.

Similar results for any other values of $m'_m s$ and $m'_s s$ can be observed for the exact closed-form BER for dual-diversity i.n.i.d. GK channels presented in this work.

## VI. CONCLUDING REMARKS

An exact closed-form expression for the BER performance of different binary modulations with dual-branch SC scheme over i.n.i.d. GK fading was derived. The analytical calculations were done utilizing a general class of special functions, specifically, the EGBMGF. In addition, this work presents numerical examples to illustrate the mathematical formulation developed in this work and to show the effect of the fading and shadowing severity and unbalance on the system performance.

ACKNOWLEDGEMENT

We would like to thank King Abdullah University of Science and Technology (KAUST), Thuwal, Makkah Province, Saudi Arabia, and Carleton University, Ottawa, Canada for providing support and resources respectively for this research work.

TABLE I

EXTENDED GENERALIZED BIVARIATE MEIJER $G$-FUNCTION (EGBMGF)

*Representation 1*: Based on [21]

$$S\begin{bmatrix} x \\ y \end{bmatrix} \equiv S\begin{bmatrix} \begin{pmatrix} p, 0 \\ A-p, B \\ q, r \\ C-q, D-r \\ k, l \\ E-k, F-l \end{pmatrix} & \begin{array}{c} (a);(b) \\ (c);(d) \\ (e);(f) \end{array} & \begin{array}{c} x \\ y \end{array} \end{bmatrix} = \begin{cases} \frac{1}{(2\pi i)^2} \int_{C_1} \int_{C_2} \frac{\prod_{j=1}^{p} \Gamma(a_j+s+t)}{\prod_{j=p+1}^{A} \Gamma(1-a_j-s-t)} \\ \times \frac{\prod_{j=1}^{q} \Gamma(1-c_j+s) \prod_{j=1}^{r} \Gamma(d_j-s)}{\prod_{j=1}^{B} \Gamma(b_j+s+t) \prod_{j=q+1}^{C} \Gamma(c_j-s)} \\ \times \frac{\prod_{j=1}^{k} \Gamma(1-e_j+t) \prod_{j=1}^{l} \Gamma(f_j-t)}{\prod_{j=r+1}^{D} \Gamma(1-d_j+s) \prod_{j=k+1}^{E} \Gamma(e_j-t)} \\ \times \frac{x^s y^t dsdt}{\prod_{j=l+1}^{F} \Gamma(1-f_j+t)}, \end{cases}$$

where $A + C < B + D$, $A + E < B + F$.

*Representation 2*: Based on [23]

$$S\begin{bmatrix} x \\ y \end{bmatrix} \equiv S\left[x, y \left| \begin{bmatrix} m_1, 0 \\ p_1, q_1 \end{bmatrix} \begin{array}{c} a_{p_1} \\ b_{q_1} \end{array} \right| \begin{pmatrix} n_2, m_2 \\ p_2, q_2 \end{pmatrix} \begin{array}{c} c_{p_2} \\ d_{q_2} \end{array} \left| \begin{pmatrix} n_3, m_3 \\ p_3, q_3 \end{pmatrix} \begin{array}{c} e_{p_3} \\ f_{q_3} \end{array} \right] \right.$$

$$\equiv G_{p_1, q_1: p_2, q_2: p_3, q_3}^{m_1, 0: n_2, m_2: n_3, m_3}\left( \begin{array}{c} a_1, \ldots, a_{p_1} \\ b_1, \ldots, b_{q_1} \end{array} \left| \begin{array}{c} c_1, \ldots, c_{p_2} \\ d_1, \ldots, d_{q_2} \end{array} \right| \begin{array}{c} e_1, \ldots, e_{p_3} \\ f_1, \ldots, f_{q_3} \end{array} \middle| x, y \right)$$

$$= \frac{1}{(2\pi i)^2} \int_{C_1} \int_{C_2} \frac{\prod_{j=1}^{m_1} \Gamma(a_j+s+t) \prod_{j=1}^{m_2} \Gamma(1-c_j+s) \prod_{j=1}^{n_2} \Gamma(d_j-s) \prod_{j=1}^{m_3} \Gamma(1-e_j+t)}{\prod_{j=m_1+1}^{p_1} \Gamma(1-a_j-s-t) \prod_{j=1}^{q_1} \Gamma(b_j+s+t) \prod_{j=m_2+1}^{p_2} \Gamma(c_j-s) \prod_{j=n_2+1}^{q_2} \Gamma(1-d_j+s)}$$

$$\times \frac{\prod_{j=1}^{n_3} \Gamma(f_j-t) x^s y^t ds dt}{\prod_{j=m_3+1}^{p_3} \Gamma(e_j-t) \prod_{j=n_3+1}^{q_3} \Gamma(1-f_j+t)},$$

where $C_1$ and $C_2$ are two suitable contours and positive integers $p_1$, $p_2$, $p_3$, $q_1$, $q_2$, $q_3$, $m_1$, $m_2$, $m_3$, $n_2$, and $n_3$ satisfy the following inequalities. $q_2 \geq 1$, $q_3 \geq 1$, $p_1 \geq 0$, $0 \leq m_1 \leq p_1$, $0 \leq m_2 \leq p_2$, $0 \leq n_2 \leq q_2$, $0 \leq m_3 \leq p_3$, $0 \leq n_3 \leq q_3$, $p_1 + p_2 \leq q_1 + q_2$, $p_1 + p_3 \leq q_1 + q_3$. The values $x = 0$ and $y = 0$ are excluded.

It may be useful to learn the relationship between both the representations shown above. Following equalities must be noted from both the above representations; $p = m_1$, $A = p_1$, $B = q_1$, $q = m_2$, $r = n_2$, $C = p_2$, $D = q_2$, $k = m_3$, $l = n_3$, $E = p_3$, and $F = q_3$.

TABLE II

EXTENDED GENERALIZED BIVARIATE MEIJER $G$-FUNCTION (EGBMGF) MATHEMATICA IMPLEMENTATION

```mathematica
(*Extended Generalized Bivariate Meijer G-Function (EGBMGF)*)
Clear All;
(*Exception*)
S::InconsistentCoeffs = "Inconsistent coefficients!";
S[{ast_, bst_}, {as_, bs_}, {at_, bt_}, {zs_, zt_}] := Module[{},

  (*Gamma product terms with only 's' as argument with other parameters *)
  Pas = Function[u, Product[Gamma[1 - as[[1, n]] + u], {n, 1, Length[as[[1]]]}]];
  Qas = Function[u, Product[Gamma[as[[2, n]] - u], {n, 1, Length[as[[2]]]}]];
  Pbs = Function[u, Product[Gamma[bs[[1, n]] - u], {n, 1, Length[bs[[1]]]}]];
  Qbs = Function[u, Product[Gamma[1 - bs[[2, n]] + u], {n, 1, Length[bs[[2]]]}]];
  Ms = Function[u, Pas[u] Pbs[u] / (Qas[u] Qbs[u])];

  (*Gamma product terms with only 't' as argument with other parameters *)
  Pat = Function[u, Product[Gamma[1 - at[[1, n]] + u], {n, 1, Length[at[[1]]]}]];
  Qat = Function[u, Product[Gamma[at[[2, n]] - u], {n, 1, Length[at[[2]]]}]];
  Pbt = Function[u, Product[Gamma[bt[[1, n]] - u], {n, 1, Length[bt[[1]]]}]];
  Qbt = Function[u, Product[Gamma[1 - bt[[2, n]] + u], {n, 1, Length[bt[[2]]]}]];
  Mt = Function[u, Pat[u] Pbt[u] / (Qat[u] Qbt[u])];

  (*Gamma product terms with 's+t' as argument with other parameters *)
  Past = Function[u, Product[Gamma[ast[[1, n]] + u], {n, 1, Length[ast[[1]]]}]];
  Qast = Function[u, Product[Gamma[1 - ast[[2, n]] - u], {n, 1, Length[ast[[2]]]}]];
  Qbst = Function[u, Product[Gamma[bst[[2, n]] + u], {n, 1, Length[bst[[2]]]}]];
  Mst = Function[u, Past[u] / (Qast[u] Qbst[u])];

  (*Countour limiters(Depends on numerator Gamma
     arguments i.e. it must be half of the least valued Gamma arguments)*)
  Rs = 1 / 4;
  Rt = 1 / 4;
  (*Assignments and Declarations*)
  Zs = zs;
  Zt = zt;
  W = 50;

  (*Final Evaluation*)
  Print["Numerical Integration:"];
  value = 1/(2 π I)^2 NIntegrate[MT[s, t] Zs^s Zt^t, {s, Rs - I W, Rs + I W}, {t, Rt - I W, Rt + I W}];

  (*Returning back the value*)
  Return[value];
];
(*End of EGBMGF*)
```

TABLE III

EXTENDED GENERALIZED BIVARIATE MEIJER *G*-FUNCTION (EGBMGF) MATHEMATICA IMPLEMENTATION USAGE EXAMPLE

Prior moving to the illustration, it may be useful to know the relationship between the representations shown in Table I and the implemented code presented in Table II. Following equalities must be noted from both the representations and the implemented code; $(a) = a_{p_1} = ast\_$, $(b) = b_{q_1} = bst\_$, $(c) = c_{p_2} = as\_$, $(d) = d_{q_2} = bs\_$, $(e) = e_{p_3} = at\_$, and $(f) = f_{q_3} = bt\_$.

An illustration of the EGBMGF implemented code usage

```
(*Testing*)
(*Declarations*)
p = 0.5; q = 1;
mm1 = 1; ms1 = 2; mm2 = 1; ms2 = 2;
Ω1 = 1; Ω2 = 1;
snr = 10^(15 / 10) (*SNR = 0 - 20 dBs*);

(*Invoking the implemented EGBMGF module*)
B =         1
    ─────────────────────────────────────────────── S[{{{p}, {}}, {{}, {}}},
    2 Gamma[p] Gamma[mm1] Gamma[ms1] Gamma[mm2] Gamma[ms2]
    {{{1}, {}}, {{mm1, ms1}, {0}}},
    {{{1}, {}}, {{mm2, ms2}, {0}}},
    { mm1 ms1   mm2 ms2  }
    { ───────, ───────   }]]
    { Ω1 snr q  Ω2 snr q  }

(*END*)
```

$0.00102393 - 7.09829 \times 10^{-16} i$

December 17, 2010                                                                                                         DRAFT14

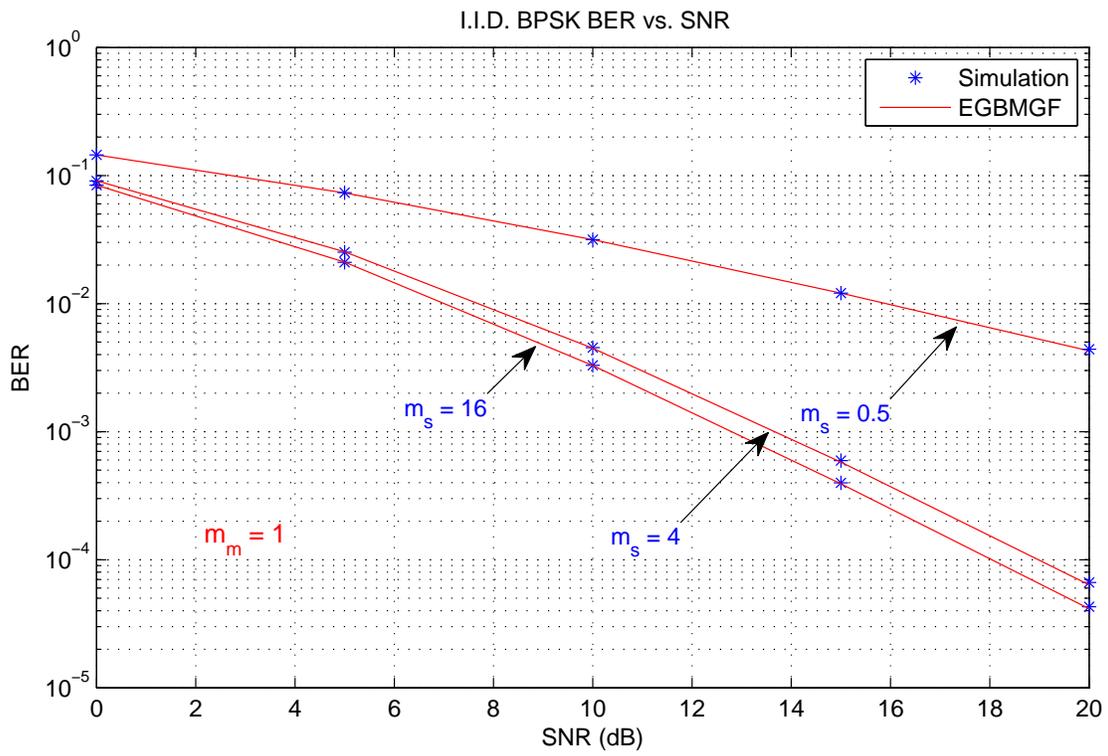

Fig. 1. I.I.D. BPSK BER for $m_m = 1$ and varying $m_s$.

16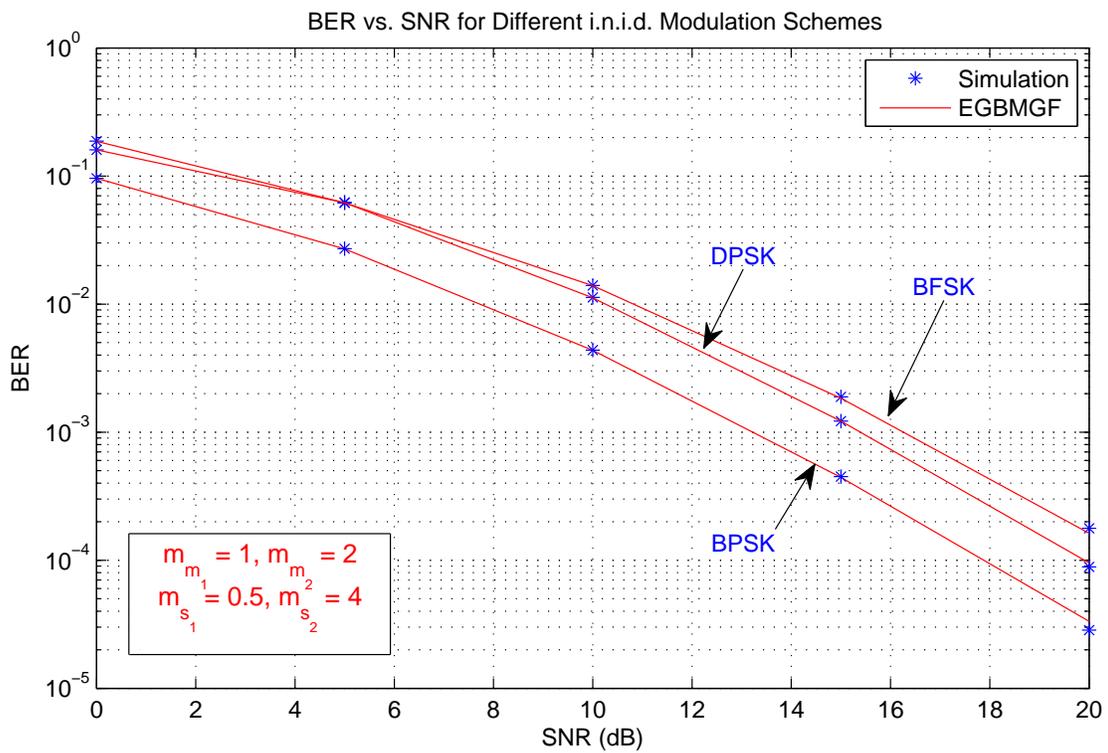

Fig. 2. BER for different modulation schemes undergoing i.n.i.d. channels with $m_{m_1} = 1$, $m_{m_2} = 2$, $m_{s_1} = 0.5$, and $m_{s_2} = 4$.

December 17, 2010  DRAFT